\documentclass[%
 reprint,
superscriptaddress,
 amsmath,amssymb,
 aps,
]{revtex4-2}

\usepackage{graphicx}
\usepackage{dcolumn}
\usepackage{bm}
\usepackage{xcolor}
\usepackage{hyperref}

\begin{document}


\title{Photon Conversion and Interaction on Chip}


\author{Jia-Yang Chen}
\altaffiliation{These authors contributed equally}
\author{Zhan Li}

\altaffiliation{These authors contributed equally}
\author{Zhaohui Ma}

\author{Chao Tang}

\author{Heng Fan}

\author{Yong Meng Sua}

\author{Yu-Ping Huang}

\email{Electronic address [1]: jchen59@stevens.edu\\Electronic address [2]: yhuang5@stevens.edu}

\affiliation {Department of Physics, Stevens Institute of Technology, 1 Castle Point Terrace, Hoboken, New Jersey, 07030, USA}
\affiliation {Center for Quantum Science and Engineering, Stevens Institute of Technology, 1 Castle Point Terrace, Hoboken, New Jersey, 07030, USA}
%


\date{\today}

\begin{abstract}
The conversion and interaction between quantum signals at a single-photon level are essential for scalable quantum photonic information technology. Using a fully-optimized, periodically-poled lithium niobate microring, we demonstrate ultra-efficient sum-frequency generation on chip. The external quantum efficiency reaches $(65\pm3)\%$ with only $(104\pm4)$ $\mu$W pump power, improving the state-of-the-art by over one order of magnitude. At the peak conversion, $3\times10^{-5}$ noise photon is  created during the cavity lifetime, which meets the requirement of quantum applications using single-photon pulses. Using pump and signal in single-photon coherent states, we directly measure the conversion probability produced by a single pump photon to be $10^{-5}$---breaking the record by 100 times---and the photon-photon coupling strength to be 9.1 MHz. Our results mark a new milestone toward quantum nonlinear optics at the ultimate single photon limit, creating new background in highly integrated photonics and quantum optical computing.
\end{abstract}

\maketitle
Unlike electrons, atoms, or any other material particles, photons do not interact with each other in vacuum. Even when mixed in optical media of the best known nonlinearities, their interaction is so weak that high optical intensities are needed to produce an appreciable effect. This inefficiency accounts for significant difficulties facing practical implementations of quantum transduction \cite{maring2018quantum},
faithful entanglement swapping \cite{sangouard2011faithful, li2019multiuser}, and heralded entanglement generation \cite{wagenknecht2010experimental, barz2010heralded}, to name a few. It also prohibits the construction of nonlinear photon-photon gates, thus forming a bottleneck for the development of scalable quantum computers at room temperature  \cite{chang2014quantum}. 
 
The recent advances in nanophotonics bring hopes to overcome this challenge, by providing tight optical confinement, strong mode overlap, and extended interaction length. Encouragingly, optical processes with improved efficiency have now been demonstrated in nanophotonic circuits made of silicon nitride \cite{li2016efficient,lu2019efficient}, aluminum nitride \cite{wang2021efficient},  gallium arsenide \cite{chang2019strong}, gallium phosphide \cite{wilson2020integrated}, aluminium gallium arsenide\cite{chang2020ultra} and lithium niobate \cite{ma2020ultrabright,chen2021efficient}. Among various candidates, thin-film lithium niobate (TFLN) on insulator has quickly arisen to a material platform of choice, due to its favorable ferro-electricity, wide optical transparency window, strong second-order nonlinearity $\chi^{(2)}$, and outstanding electro-optical responses. Compared with third-order nonlinear ($\chi^{(3)}$) materials, TFLN is centro-asymmetric and possesses exceptionally large second-order nonlinearity ($\chi^{(2)}$) to produce orders of magnitude stronger effects, as desirable to optical nonlinearities at a single photon level. Thus far, a variety of TFLN microresonators have been demonstrated with impressive  nonlinearities \cite{lu2019periodically,chen2019ultra,Lu:20,ma2020ultrabright,chen2021efficient,gao2021broadband}. However, their performance has been capped by the use of relatively small $\chi^{(2)}$ susceptibilities (e.g., $d_{31}\sim$ 4.7 pm/V) \cite{lu2019periodically}, poor mode overlapping \cite{chen2019ultra,gao2021broadband}, and/or low photon-extraction efficiency \cite{lu2019periodically,chen2019ultra,Lu:20,ma2020ultrabright}. For photon conversion and interaction, while their high efficiency has been predicted from, e.g., second-harmonic generation (SHG) \cite{lu2019periodically,chen2019ultra,Lu:20} or parametric downconversion experiments \cite{ma2020ultrabright}, there has been no direct demonstration.       

\begin{figure*}[ht]
\centering
\includegraphics[width= 6.5 in]{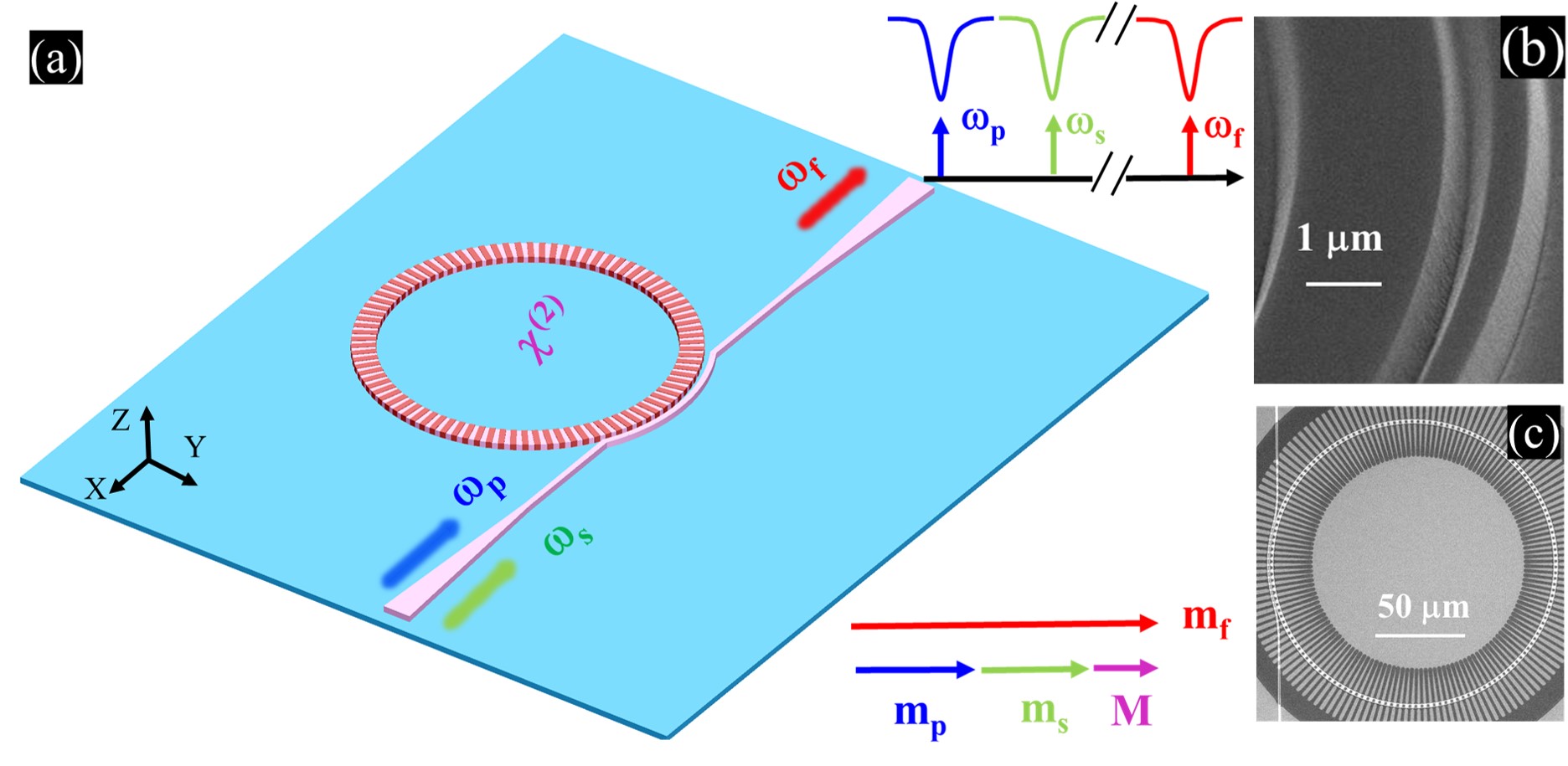}
\caption{Integrated TFLN circuits for photon conversion and interaction. (a) Schematic of the Z-cut periodically poled microring, where the pump of $\omega_p$ and signal of $\omega_s$ couple into the microring and generate the SF light of $\omega_{f}$ via a $\chi^{(2)}$ process. A pulley coupler is designed for over coupling all light waves, for high photon extraction efficiency. Insets illustrate triply resonant and quasi-phase matching conditions. (b) and (c) are the SEM images of the etched pulley coupler before poling process and the periodic poled microring after removing the poling electrodes, respectively.}
\label{fig1}
\end{figure*}

Here, we present a TFLN resonator that overcomes all aforementioned shortcomings and delivers its promised high efficiency for photon conversion and interaction. As illustrated in Fig.\ref{fig1}(a), it is an overly coupled, triply resonant, periodically poled lithium niobate (PPLN) microring resonator for sum-frequency generation (SFG). All interacting light waves are in the low-loss fundamental modes with nearly perfect overlap and interact through TFLN's largest $\chi^{(2)}$ susceptibility tensor element (e.g., $d_{33}\sim$ 27 pm/V). It achieves an impressive photon-photon coupling strength of $g$ = 9.1 MHz (angular frequency). Crucially, by strongly overcoupling the cavity to minimize the extraction loss of the sum-frequency (SF) photons, we demonstrate photon conversion at a record-high external efficiency of $65\%$ with only about 100 $\mu$W pump power, marking orders of magnitude improvement over the state of the art across all existing photonic platforms; see Table \ref{table1}. At the peak conversion, the on-chip noise photon flux is only $3\times10^{-5}$ photons per 100-ps cavity lifetime, despite small-detuning pumping. This ultrahigh external efficiency yet low noise create new opportunities in various applications like quantum frequency conversion, optical squeezing, and phase sensitive amplification. 

\begin{table*}[ht]
\centering
\begin{tabular}{ccccccc}
Material Platform & Structure & Nonlinear Process &~~~$Q_l$ ($\times10^5$)~~~& $ \eta_\mathrm{con}$ & ~~Pump Power ~~ & Ref.\\\hline
Si$_3$N$_4$ & microring  & FWM-BS & 1.5/1.5/2.4           & 60\% &50 \& 8 mW  &\cite{li2016efficient}\\
Si$_3$N$_4$ & microring  & DFWM &   2.8/3.0/1.2   & 13\% &0.33 mW  &\cite{lu2019efficient}\\
AlN & microring  & SFG &  3.0/3.0/1.4 & 42\% &35 mW &\cite{wang2021efficient}\\
PPLN & microring & SHG&           1.3/3.0          &10\% &0.3 mW  & \cite{ma2020ultrabright}\\
PPLN & microring & SHG&           1.5/0.6          & 26\% &1.8 mW  &\cite{chen2021efficient}\\
PPLN & millimeter disk  &SHG&         120/80          & 18\% &9 mW  &\cite{ilchenko2004nonlinear}\\
PPLN & microring & SFG &           1.3/1.3/2.6   & 65\% & 0.1 mW  &this work\\
\hline
\end{tabular}
\caption{The state-of-the-art frequency conversion in $\chi^{(2)}$ and $\chi^{(3)}$ cavities. $\eta_\mathrm{con}$: conversion efficiency by photon number;  FWM-BS: four-wave mixing Bragg scattering; DFWM: degenerate four-wave mixing. $Q_l$ lists the loaded quality factors for the pump, signal and SF waves in the case of FWM-BS, DFWM, and SFG,  and the pump and second-harmonic waves in the case of SHG. This table only includes those whose conversion efficiency is over $10\%$. Note: a recent arxiv preprint reported SHG of $\eta_{\mathrm{con}} = 33\%$ \cite{gao2021broadband}. According to its reported normalized efficiency of $602\%$/mW, the required pump power shall be $P_p = 16~\frac{66\%}{602\%/\mathrm{mW}} = 1.75~\mathrm{mW}$ \cite{lu2019periodically,chen2021efficient}, which is over 16 times larger than its claimed value. We have not included it in this table due to this apparent inconsistence.}
\label{table1}
\end{table*}

Beside the large coupling strength,
our device's high cavity quality for all interacting waves provides extended interaction length to boost optical nonlinearities towards the single photon regime. To assess this prospect, we further perform photon interaction between two single-photon signals in weak coherent states. Our measurement directly shows that with only one pump photon in the microring, the external quantum efficiency for signal photon conversion is $\sim 10^{-5}$, while the best efficiency reported hitherto is $10^{-7}$ \cite{li2017nonlinear}. The internal Rabi oscillation angle produced by the pump photon is 0.01, which can be improved to approach unity by further reducing the cavity loss. 

As a whole, our results constitute a new milestone along the long pursuit of nonlinear optics in its ultimate quantum limit, where a single photon is enough to produce significant nonlinear effects. While there is still a good journey to take before hitting the finish line, the fact that we can attain the theoretical performance of this device is critical for us to take steps forward. By further improving the cavity quality, strong photon-photon interaction is within sight. Based on the current nonlinear parameters, a cavity qualify factor of $10^8$---which has been demonstrated in lithium niobate microdisks \cite{gao2021broadband}---will enable C-NOT gate between single photons. Meanwhile, the demonstrated photon conversion and interaction efficiency can already elevate the performance of nonlinear-optical devices for heralded entanglement generation, faithful entanglement swapping, and so on. 

\textbf{Device design:}  In $\chi^{(2)}$ cavity, the effective Hamiltonian describing SFG between photons in their single modes is    
\begin{equation}
\hat{H}_\textrm{eff}  = \hbar\mathnormal{g}(\hat{a}_p\hat{a}_s\hat{a}_{f}^\dagger+\hat{a}_p^\dagger\hat{a}_s^\dagger\hat{a}_{f}), \label{eq1}\\
\end{equation}
where $\{\hat{a}_j\}$ are the annihilation operators with $j = p,s,f$ standing for the pump, signal and SF light, respectively, each with angular frequency $\omega_j$. ${g}$ is the photon-photon coupling strength, which can be interpreted as the effective Rabi frequency produced by a pump photon (see Supplementary Material\ref{supp1}). It is given by
\begin{equation}
g \propto \frac{d_\textrm{eff}\xi}{\sqrt{V_\textrm{eff}}}\times\delta(m_{f}-m_{p}-m_{s}-M)\label{eq2}
\end{equation}
where $d_\textrm{eff}$ is the effective nonlinear susceptibility. $\xi$ is the mode overlapping factor. $V_\textrm{eff}$ is the effective mode volume. $m_j$ is the azimuthal order of the cavity modes, and $M$ is the azimuthal poling grating number, so that $\delta(m_{f}-m_{p}-m_{s}-M)$ accounts for quasi-phase matching (QPM) by periodic poling. 

In this work, we use a microring with a radius of 80 $\mu$m and a cross-section of 600 nm in height and 1700 nm in top-width. The pump and signal are both in the telecom C-band and  their SF is in the visible band, chosen so with repeater-based quantum communications in mind. To maximize $g$, all three waves are in the fundamental quasi-transverse-magnetic (quasi-TM) modes and interact through TFLN's largest nonlinear tensor $d_{33}$ with over 90\% mode overlap. As shown in Fig.\ref{fig1}, concentric periodic poling is applied to the microring for QPM. To ensure triple resonances for all waves, fine temperature tuning ($\sim$ 0.01 $^\circ$C) is applied to compensate for any resonant mismatch due to any fabrication error. 

For photon conversion and interaction, the device figure of merit is the external quantum efficiency (QE) defined as
\begin{equation}
    \eta_\mathrm{QE} = \frac{N_{f}}{N_s},\label{eq3}
\end{equation}
where $N_s$ is the number of input signal photons to the cavity and $N_f$ is that of converted SF photons at the cavity output (thus accounting for any internal cavity loss). This is in contrast to previous demonstrations, where critical coupling was adopted to maximize the intracavity optical power for high conversion efficiency \cite{lu2019periodically,chen2019ultra,Lu:20,ma2020ultrabright}. However, most of the input power and about half of the converted photons are lost inside the cavity, rendering a rather low QE while prohibiting cascaded operations. For practical applications, one instead needs to over-couple the cavity so that the photons can be extracted out before significantly lost in the cavity. 

Under QPM and triple resonance (see Supplementary Material\ref{supp1}), the maximum QE is given by \cite{ breunig2016three}:
\begin{align}
  \eta_\mathrm{QE}^\mathrm{max} \approx \frac{Q_{s,l}}{Q_{s,c}} \frac{Q_{f,l}}{Q_{f,c}},\label{eq4}
\end{align}
where $Q_{j,o}$ is the quality factor with $o = c,l$ denoting the coupling and loaded Q, respectively. It is reached with an optimal pump power 
\begin{align}
  P_{p}^\mathrm{opt} \approx 8~\frac{\eta_\mathrm{QE}^\mathrm{max}}{\eta_\mathrm{tran}^\mathrm{nor}},\label{eq5}
\end{align}
where $\eta_\mathrm{tran}^\mathrm{nor} = P_{f}/(P_{s}P_{p})$ is the normalized power transduction efficiency with $P_j$ the optical power of the $j$-th wave. In our case, $\eta_\mathrm{QE}^\mathrm{max}\approx65\%$ and  $\eta_\mathrm{tran}^\mathrm{nor}\approx4.5\%/ \mu$W, so that the optimal power is around 115 $\mu$W. 

We use a pulley coupler in optimized dimensions to achieve proper over-coupling for both the signal and SF modes. This is done by first determining its top width by requiring $n_{p} R_{p} = n_{r} R_{r} $ \cite{hosseini2010systematic}, where $n_{p,r}$ are the effective refractive indices of the pulley waveguide and microring modes for the sum-frequency wave, and $R_{p}$ and $R_{r}$ denote their radii. For the SF wave, the waveguide-microring coupling strength is proportional to the length of the pulley coupler, while inversely proportional to their gap. For the signal mode, in contrast, the coupling strength varies as $\mathrm{sinc}(\Delta\Phi)$, where $\Delta\Phi$ is the coupling phase mismatch. This allows to carefully design the dimensions to create the desirable over-coupling for both signal and sum-frequency waves. 
\begin{figure*}[ht]
\centering
\includegraphics[width= 6.5 in]{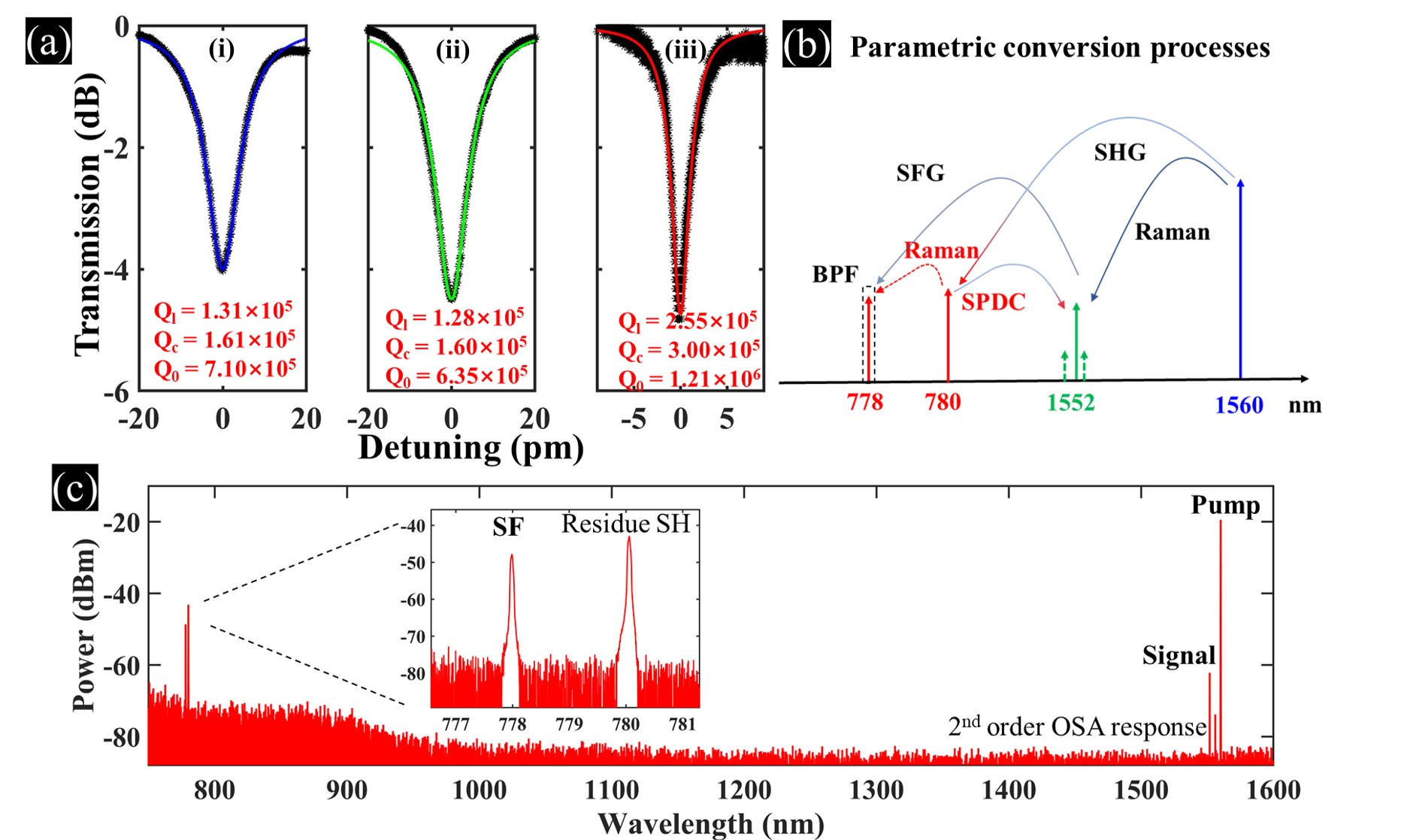}
\caption{(a) Optical spectra of the interacting TM$_{00}$ cavity modes at (i) 1560.15 nm, (ii) 1551.85 nm,   and (iii) 778.00 nm, respectively. (b) Illustration of possible parametric conversion processes in the microring. (c) Original optical spectrum before any optical filtering when strong pump and weak signal are applied. In high conversion region, signal starts to be depleted. Insert gives the zoom-in spectrum around the SF band. After correcting for the coupling loss, the on-chip power of pump, signal and SF waves are -13.8 dBm, -44.3 dBm and -44.0 dBm, respectively.}
\label{fig2}
\end{figure*}

\textbf{Device characterization:} 
The details of our device fabrication and experiment setup are presented in Method and Supplementary \ref{supp2}. The fiber-chip-fiber coupling losses are measured to be 8 $\pm$ 0.15 dB at 1556 nm and 9.5 $\pm$ 0.2 dB at 778 nm, respectively. To find phase matching, we first search for strong SHG across the C-band by sweeping an infrared laser while optimizing the chip's temperature. This gives several cavity modes around 1556 nm, based on which we select a set of cavity modes at  1560.15 nm, 1551.85 nm, and 778.00 nm, considering the over-coupling requirement and limited by our visible band-pass filter (BPF, bandwidth $\sim$3 nm, 760 to 780 nm).  To reduce the Raman background, we designate the 1560.15 nm mode to the pump. For this set, $m_{p} = 602$, $m_{s} = 606$, $m_{f} = 1357$, so that $M = 149$. The loaded quality factors $Q_{j,l}$ are measured for each modes, while the coupling $Q_{j,c}$ and intrinsic  $Q_{j,0}$ factors are calculated by fitting the resonance spectra, as shown in Fig.~\ref{fig2} (a). 
Those Q's, according to Eq.~(\ref{eq2}), give the highest possibl external  quantum efficiency of $\eta_\mathrm{QE}^\mathrm{max} \approx 65\%$. For an even higher efficiency, the cavity needs to be further overcoupled. 

\textbf{Low-noise Frequency Conversion: } 
We perform SFG using the setup detailed in Fig.~\ref{fig6}. We first couple a strong pump at 1560 nm and a weak signal at 1552 nm into the microring, and fine tune the microring temperature and the laser wavelengths to verify QPM and triple resonance. The resulting spectrum, measured without any filtering thus manifesting all possible nonlinear processes, is shown in Fig.~\ref{fig2}(c). It exhibits a clean profile with low baseline, except for a residual second-harmonic peak at $\sim$780 nm by the strong pump. This peak is significant only in the high conversion regime (i.e.,$>$50$\%$) and can be conveniently filtered out. In this experiment, we reject it using a $\sim$3 nm bandpass filter centered at 778 nm. The otherwise low background over the entire spectral range shows that all other competing processes are well suppressed, as desirable for quantum applications. 

\begin{figure}[ht]
\centering
\includegraphics[width=3.2 in]{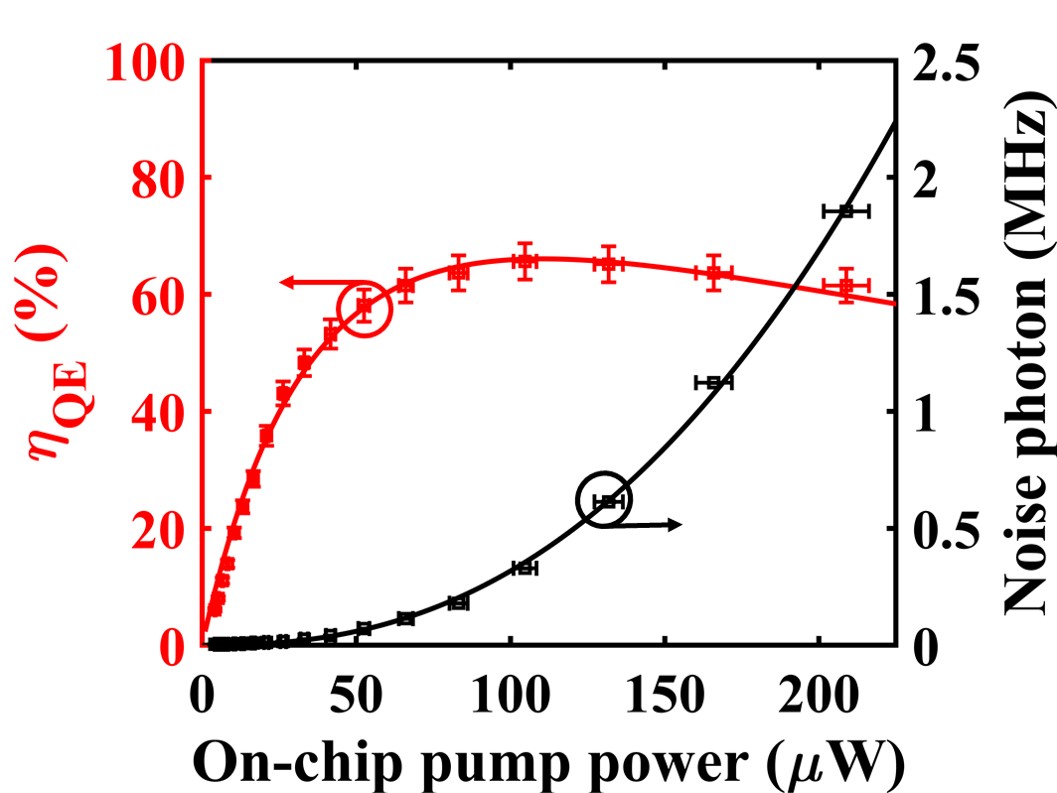}
\caption{SFG efficiency  $\eta_\mathrm{QE}$ (left-axis, red square) and the generated noise photon flux $N_\mathrm{noise}$(right-axis, black square) plotted against the on-chip pump power. $\eta_\mathrm{QE}$ = 65$\%$ is obtained at the pump power around 100 $\mu$W. Solid red curve represents the prediction of the coupled mode equations (see Supplementary Material\ref{supp1}) using the actual parameters of the microring with only one fitting parameter: $g$ = 8.2 MHz. Solid black line is fitted to the pump power $P$ as  $N_\mathrm{noise}\sim P^{2.4}$. The error bars in left-axis and x-axis are estimated according to coupling instability. The error bar of noise photon in right-axis is estimated by uncertainty assuming Poissonian photon counting statistics.}
\label{fig3}
\end{figure}

In this experiment, the signal power is fixed to be about 37 nW on chip, while the pump power is gradually increased. During the measurement, we only need to slightly optimize the temperature within 45 $\pm$ 0.3 $^\circ$C and tune the wavelengths of both lasers within $\pm$ 20 pm to compensate for slight phase mismatch and resonance drift caused by thermo-optical and photorefractive effects \cite{chen2021efficient}. The quantum efficiency as a function of the on-chip pump power is shown in Fig.~\ref{fig3}. Thanks to the over-coupling and nearly ideal poling, $(65\pm3)\%$ quantum efficiency is achieved with at $(104\pm4)$ $\mu$W pump power. Taking into account all insertion losses both on and off chip, this corresponds to about $9\%$ total efficiency. In the low conversion region, the normalized power transduction efficiency is fitted to be about 4.5$\%$ $\mu$W$^{-1}$. By fitting the experimental data with the steady-state solution to Eqs.(S1-S3), as shown in Fig.~\ref{fig3}, the photon-photon coupling coefficient $g$ is determined to be 8.2 MHz. 

The above results speaks to the ultrahigh efficiency. For quantum frequency conversion, it is critical that no significant in-band noise is injected during the conversion. As illustrated in Fig.~\ref{fig2}(b), there are multiple processes that can produce in-band noise, such as Raman scattering from strong classical pump to the signal band followed by SFG, Raman scattering by the pump's residue SH light, and spontaneous parametric down-conversion (SPDC) followed by SHG and SFG. To quantify their total contributions, we measure the noise photon flux generated in the SF band when only the pump is applied. To ensure total rejection of any out-band noise, the SF photons are passed through additional free-space filters (see Supplementary Material \ref{supp2}), before detected by a silicon-based single-photon detector (Si-SPD, quantum efficiency: 50$\%$, dark count: 250 Hz). The results are plotted in Fig.~\ref{fig3}, where the increase of the noise photon flux is between quadratic and cubic with the pump power. This result indicates that besides Raman scattering, the cascaded SPDC and SFG process also presents, similar to what we observed in a PPLN nanowaveguide \cite{fan2021photon}. At the $65\%$ peak QE, the on-chip noise photon flux is 0.3 MHz under continuous-wave pumping. If using $\sim$100 ps pulses that match the cavity lifetime, the noise photon per pulse is $3\times10^{-5}$, which is low especially given the small detuning between the pump and signal that are both in the telecom C-band. This noise level can be substantially lowered by, for example, further detuning the pump from the signal \cite{singh2019quantum}.  

\begin{figure}[ht]
\centering
\includegraphics[width=3 in]{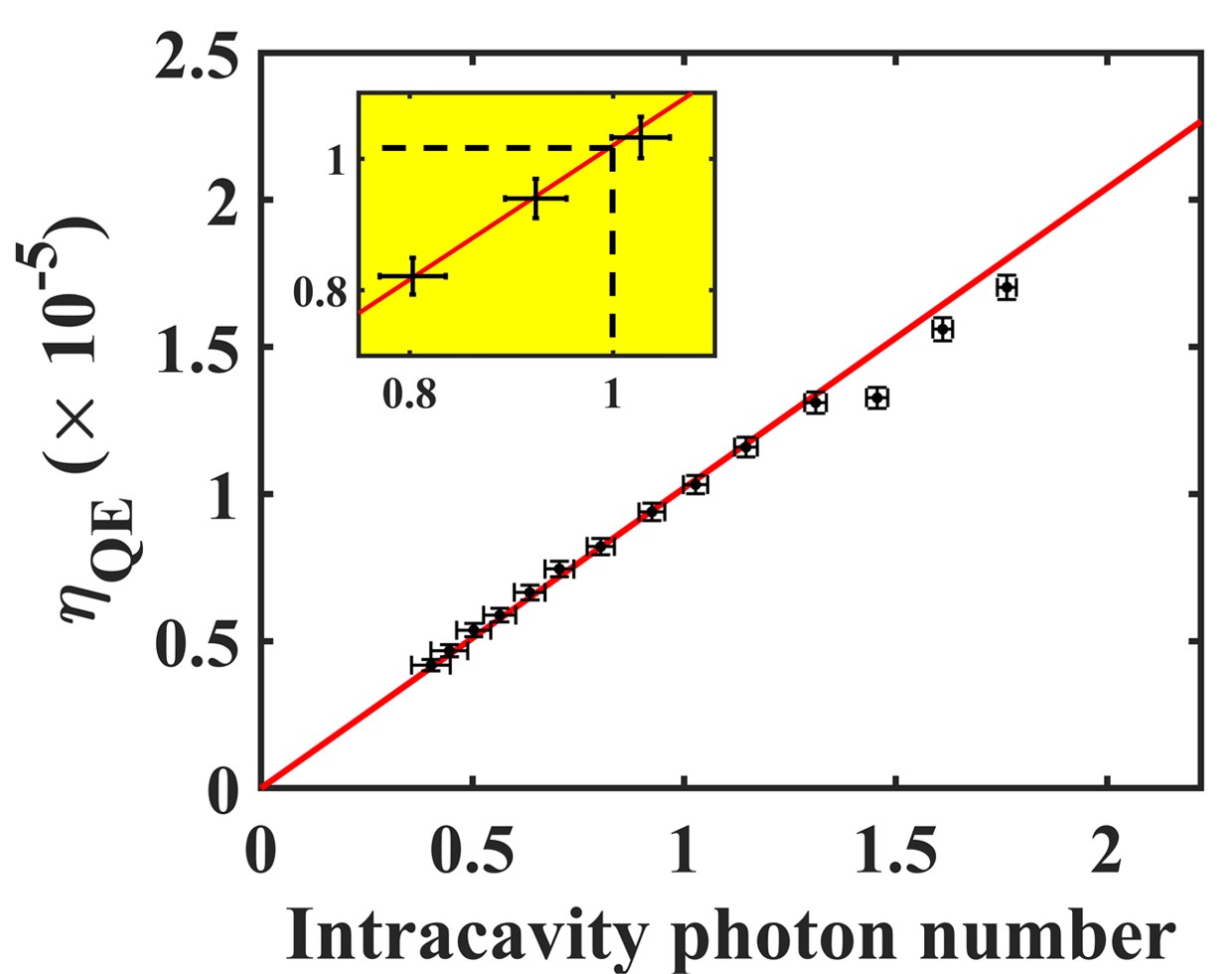}
\caption{Quantum efficiency $\eta_\mathrm{QE}$ versus the intracavity mean pump photon number. Detector dark counts of 250 Hz have been subtracted, and the coupling loss of 4.75 dB and detector efficiency of 50$\%$ have been accounted for. Inset is an zoom-in to shown a quantum effciency of $\eta_\mathrm{QE}=10^{-5}$ is achieved at one intracavity pump photon (corresponding to photon flux $N_{p} = 2.8$ GHz), where the intracavity signal photon number is fixed at 0.25 (corresponding to photon flux $N_{s} = 0.7$ GHz). Solid red line represents the simulated results using the actual parameters of the microring with only one free parameter $g$ = 9.1 MHz, fitted with the coupled-mode equations (see Supplementary Material\ref{supp1}). All error bars are estimated assuming Poissonian photon counting statistics.}
\label{fig4}
\end{figure}

\begin{figure*}[ht]
\centering
\includegraphics[width= 6.5in]{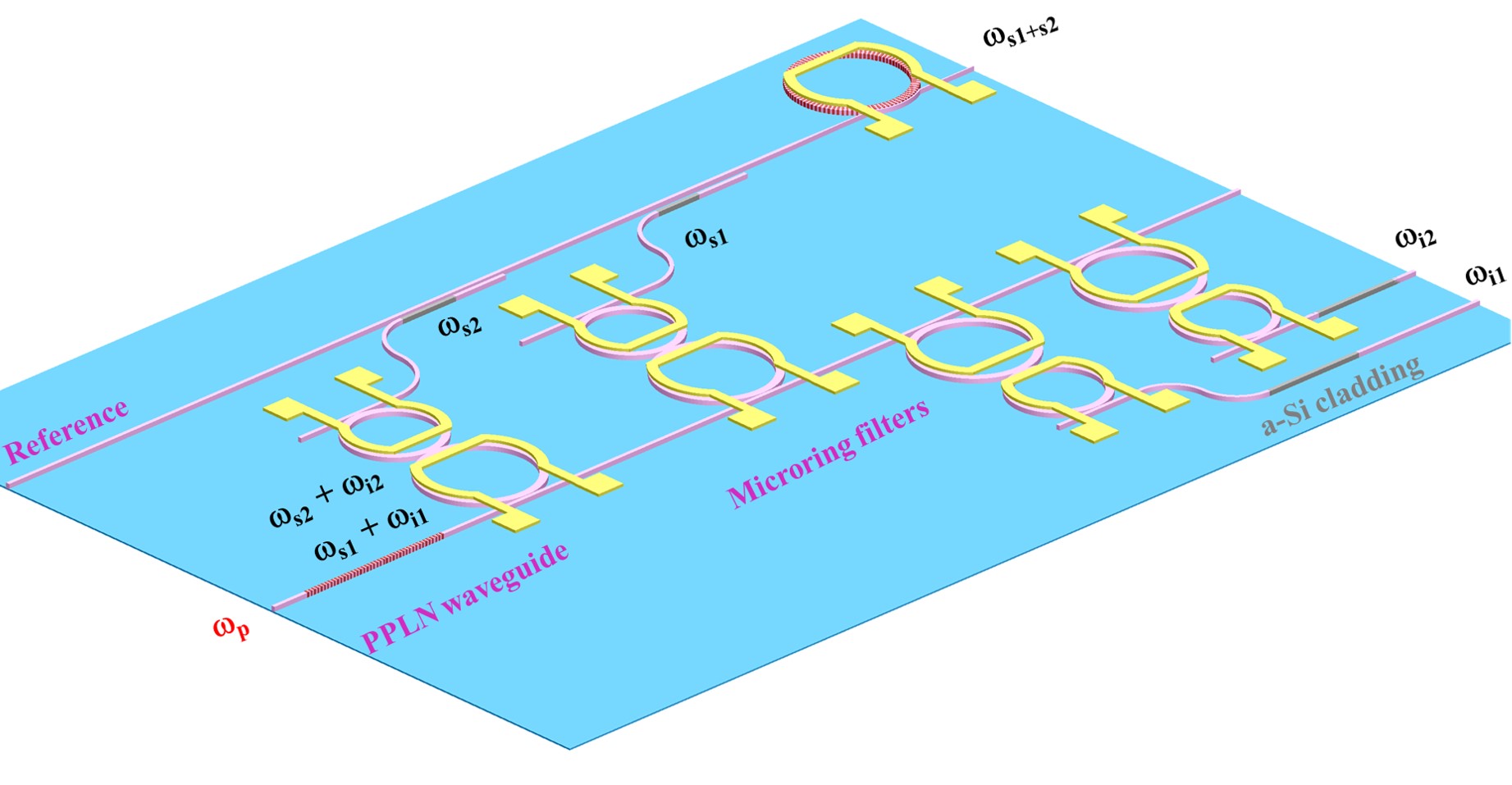}
\caption{Illustration of a TFLN integrated photonic circuit for heralded entanglement generation.}
\label{fig5}
\end{figure*}

\textbf{Interaction Between Photon-level Coherent States:}  Strong interaction between single photons is of great values for both fundamental optics studies and quantum applications such as faithful quantum entanglement swapping \cite{sangouard2011faithful, li2019multiuser}, heralded entanglement source \cite{wagenknecht2010experimental, barz2010heralded}, and device-independent quantum key distribution \cite{vazirani2019fully}. Here, we characterize the device responses in the single photon regime by attenuating the pump and signal to weak coherent states at single photon levels. Their photon fluxes are carefully monitored by two superconducting single-photon detectors (SNSPDs, see Supplementary Material \ref{supp2}). 

To maximize the detection efficiency for the SF photons, we remove the free-space filtering system---used in the last section to reject background noise created by the strong pump---and directly couple the SF photons from the chip to the Si-SPD via a lensed fiber. To ensure that the pump photons do not generate background counts (that could contribute to the over-estimation of the single-photon nonlinearity), we block the signal and verify that even at the highest on-chip flux ($\sim$ 5 GHz), the photon counts remain at its dark count level ($\sim$ 250 Hz). Here, the total detection efficiency for the sum-frequency photons is $\eta = \eta_d\eta_c \approx 17\%$, with detector efficiency $\eta_d\sim50\%$ and coupling efficiency $\eta_c\sim33.5\%$. 

The measurement results are shown in Fig.~\ref{fig4}, where we observe linear dependency of the quantum efficiency on the intracavity pump photon number. The on-chip quantum efficiency is about $10^{-5}$ when there is one pump photon on average in the cavity, after correcting for the coupling loss and finite detector efficiency. By fitting with the coupling-mode equations, we extract the photon-photon coupling coefficient $g$ to be 9.1 MHz, compared to 8.2 MHz as from the classical measurement. This slightly higher value may come from better QPM or triple resonance for single-photon light waves, due to the absence of thermo-optical and photorefrative effects.

The above quantum efficiency gives the probability of a signal photon at the cavity input being converted to its SF and appear at the cavity output. It has accounted for the signal coupling loss and the SF extraction loss, thus constituting a direct measure of the device figure of merit that dictates its performance in quantum applications. Another measure, of less practical relevance but nonetheless describing an intrinsic property, is the internal Rabi rotation angle $\theta=2g Q_{p,i}/\omega_p$. That's, Eq.~(\ref{eq1}) can be interpreted as the pump induced Rabi oscillation between signal and SF photons. $\theta$ then gives how much Rabi rotation can a pump photon induce during it is lost in the cavity. In our case, $g=9.1\times10^6$ and $Q_{p,i}=7.1\times 10^5$, so that $\theta=0.01$, which is already appreciable from a fundamental standpoint. Increasing $\theta$ to $\pi/2$ would require improving $Q_{p,i}$ to $10^8$, which has been demonstrated in polished TFLN microdisks \cite{gao2021broadband}.

\textbf{Heralded Entanglement Generation:}
The demonstrated photon-level nonlinearity implies unprecedented performance in quantum information science and technology. In particular, because of lithium niobate's exceptional optical properties in multiple aspects, the demonstrated photon conversion and interaction are ready to be integrated with other passive and active elements on the same chip, such as PPLN wavegudies \cite{Chen:19,chen2020efficient}, electro-optical modulators \cite{jin2019high}, frequency comb sources \cite{he2019self,wang2019monolithic,zhang2019broadband,gong2020near}, and microring filters \cite{jiuyi}, to create functional quantum devices of practical impacts \cite{boes2018status}. Compared with the existing table-top or assembled systems, the detrimental insertion loss will be eliminated, and the mechanical and optical stabilities are expected to be exceptional.  

As an example, in Fig.~\ref{fig5} we present the design of an integrated chip for heralded entanglement generation, following the scheme in \cite{sangouard2011faithful}. It consists of a PPLN nanowaveguide to create photon pairs by SPDC, a series of thermally-tuned microrings for optical filtering, directional couplers for photon combination, and a PPLN microring for photon conversion. The process starts with passing a visible pump through a PPLN nanowaveguide to generate via SPDC photon pairs simultaneously over multiple wavelength channels, $\omega_{s1} + \omega_{i1}$, $\omega_{s2}+\omega_{i2}$,..., and $\omega_{sn}+\omega_{in}$ \cite{Chen:19,chen2020efficient}. The signal and idler photons will each be picked and separated into different arms by using coupled add-drop filters for high extinction while eliminating free spectrum ambiguity. In each arm, amorphous-silicon over-cladding will be applied to further reject any residue pump in the visible band. The signal photons will be recombined via cascaded directional couplers and sent into a PPLN microring, where they are interact to create a SF photon. The existence of a photon pair in the idler channels will be  heralded upon the detection of the SF photon. To create entanglement in time bins, the SF photon needs to pass through a Franson interferometer and be detected in a superposition time-bin state \cite{sangouard2011faithful}. Unlike schemes using SPDC and linear optical Bell state measurement, where the success probability is fundamentally capped at 50\% \cite{zhang2008demonstration}, this scheme is deterministic in that upon heralding, the entangled photon pair exists with nearly certainty. In this design, with 12~pm net filtering bandwidth and 100 ps pump pulses, one can drive the SPDC at 1\% photon pair production rate per pulse for each channel. The heralding rate can approach 10 Hz for the demonstrated $10^{-5}$ photon conversion, which would correspond to orders of magnitude improvement over previous demonstrations \cite{wagenknecht2010experimental, barz2010heralded}·

\textbf{Discussion:} 
We have demonstrated photon conversion and interaction with record high efficiency and low noise in a Z-cut, periodically poled microring on thin-film lithium niobate. Combining nearly perfect QPM, tight mode confinement, high cavity quality, and efficient photon extraction, we achieved 65\% wavelength transduction with only about 100 microwatt pump power, advancing the state of the art by large. Despite a small detuning between the signal and pump, at the peak conversion only $3\times10^{-5}$ noise photons are created over the cavity lifetime, thanks to the deep single-mode condition and suppression of side processes. The same device allowed nonlinear interaction between two single-photon level coherent states, where the photon-photon coupling strength reaches 9.1 MHz, and a single photon can produce 0.01 internal Rabi rotation angle. The external quantum efficiency is directly measured to be about $10^{-5}$, compared with the best reported efficiency of $10^{-7}$. Our results mark new milestones towards quantum nonlinear optics in the single photon regime, with broad implications in areas of fundamental studies and applied quantum information technology. Combining with other favorable optical proprieties of thin film lithium niobate, a superior platform of photonic integrated circuits is within sight for scalable quantum applications.  

\newpage
\section*{Methods}\label{Mtd}
The entire device is fabricated on a magnesium-doped Z-cut LNOI wafer (NANOLN Inc.), with a 600-nm thick LN thin film bonded on 2 $\mu$m silicon dioxide layer above a silicon substrate. First, the microring and waveguide structure are defined using hydrogen silsesquioxane (HSQ, Fox-16) by electron beam lithography. The top width and the radius of the microring are 1.7 $\mu$m and 80 $\mu$m, respectively. Then, ICP Argon milling is applied to shallowly etch the structures, where 430-nm thick LN is etched with 170-nm LN remaining, and the sidewall angle is approximately 64$^\circ$. The optimized pulley coupler, shown in Fig.~\ref{fig1}(b) with the pulley top width of $w_{pulley}$ = 400 nm, the gap of $g_{pulley}$ = 650 nm, and the length of $L_{pulley}$ = 40 $\mu$m, is created to increase the ring-bus waveguide coupling to attain simultaneously over-coupling condition for both the visible and IR lightwaves. Then, a concentric periodically-poled region with period of $\Lambda = 3.37$ $\mu$m given by $\Lambda = 2\pi R/M$, M = 149 and near 50$\%$ duty cycle (see Fig.~\ref{fig1}(c)) is created via several 1-ms and 450-V electrical pulses using a similar process described in \cite{chen2020efficient}. After removing the poling electrodes, the chip is coated with 1.5 $\mu$m silicon dioxide. Finally the chip is diced and polished.

\section*{Acknowledgements}
The research was supported in part by National Science Foundation (Award \#1641094 \& \#1842680 \& \#1806523) and National Aeronautics and Space Administration (Grant Number 80NSSC19K1618). Device fabrication was performed in Nanofabrication Facility at Advanced Science Research Center (ASRC), City University of New York (CUNY).

\section*{Author contributions statement}
J. C. and Y. H. conceived the experiments. J. C. and Z. L. fabricated the device and conducted the experiments. Z. M. and C. T. involved in the fabrication. H. F. and Y. S. involved in the quantum measurement. J. C. performed the numerical simulations. Y. H. supervised the project. All authors contributed to write and review the manuscript. 


\section*{Competing interests}
The authors declare no competing interests.

\section*{Additional information}
Correspondence and requests for materials should be addressed to J.C and Y.H.

\clearpage
\widetext
\begin{center}
\textbf{\large Supplemental Materials: Photon conversion and interaction on chip}
\end{center}
\setcounter{equation}{0}
\setcounter{figure}{0}
\setcounter{table}{0}
\setcounter{page}{1}
\makeatletter
\renewcommand{\theequation}{S\arabic{equation}}
\renewcommand{\thefigure}{S\arabic{figure}}
\renewcommand{\thetable}{S\arabic{table}}
\renewcommand{\bibnumfmt}[1]{[S#1]}
\renewcommand{\citenumfont}[1]{S#1}

\subsection{Coupled-mode theory}\label{supp1}
The dynamic of sum-frequency generation (SFG) inside the $\chi^{(2)}$ cavity, neglecting Rayleigh backscattering and assuming the single-mode condition, is governed by the following coupled mode equations:
\begin{align}
\frac{da_p}{dt}    &=(i\delta_p-\frac{\kappa_{p,t}}{2})a_p+ig^*a^*_sa_{f}+i\sqrt{\kappa_{p,c}}F_p,\\
\frac{da_s}{dt}    &=(i\delta_s-\frac{\kappa_{s,t}}{2})a_s+ig^*a^*_pa_{f}+i\sqrt{\kappa_{s,c}}F_s,\\ 
\frac{da_{f}}{dt} &=(i(\delta_{p}+\delta_{s}+\Delta\omega)-\frac{\kappa_{f,t}}{2})a_{f}+iga_pa_s,
\end{align}
where $\kappa_{j,o} = \omega_j/Q_{j,o}$ are the cavity dissipation rates, $Q_{j,o}$ are the quality factors of cavity mode, $\omega_j$ is the angular frequency, with $j=p,s,f$ for the pump, signal, sum-frequency modes, respectively, and $o=0, c, t$ denoting intrinsic, coupling and total dissipation rates with $\kappa_{t} = \kappa_{0} +\kappa_{c}$, $\delta_j = \omega_j-\omega_{j,0}$ is the laser-cavity detuning and $\omega_{j,0}$ is the cavity resonance, $ F_j = \sqrt{N_j} $ with $N_j$ is the photon number. Because of the energy conservation in parametric conversion process, $\omega_{f}=\omega_{p}+\omega_{s} $ permits $\delta_{p}+\omega_{p,0}+\delta_{s}+\omega_{s,0}=\delta_{f}+\omega_{f,0}$ thus $\delta_{f} = \delta_{p}+\delta_{s}+\Delta\omega $ where  $\Delta\omega = \omega_{p,0}+\omega_{s,0}-\omega_{f,0}\approx0$, when it is a triply-resonant cavity. The photon-photon coupling coefficient $g$:
\begin{align}
g = \sqrt{\frac{\hbar\omega_p\omega_s\omega_{f}}{2\epsilon_0 \epsilon_p\epsilon_s\epsilon_{f}}} \frac{\frac{2}{\pi} d_{eff}\xi}{
\sqrt{A_{eff}2\pi R}}\times\delta(m_{f}-m_{p}-m_{s}-M),\\\label{}
\xi = \frac{\iint E^*_{f}E_pE_s dxdy}{(\iint | E_p |^2  E_p dxdy \iint | E_s |^2 E_s dxdy \iint | E_{f} |^2 E_{f} dxdy)^{1/3}},\label{}
\end{align}
where $\epsilon_0$ is the vacuum permittivity, $\epsilon_j = n^2_j$ are the relative permittivity with $n_j$ giving the effective refractive indices, $d_{eff}$ is the effective nonlinear susceptibility, $\xi$ is the mode overlapping factor and $V_{eff}$ is the effective mode volume, $m_j$ are the azimuthal order of the cavity modes, M is the azimuthal poling grating number. For quasi-phase matched case, $ m_{f}-m_{p}-m_{s}-M = 0$.

At steady state, the output filed of SF is defined as $b_{f} = i\sqrt{\kappa_{f,c}}a_{f} $ and the quantum efficiency of photon conversion is given by $\eta_\mathrm{QE} = N_{f}/N_s =\lvert b_{f} \rvert^2/N_s$. The intracavity pump and signal photon number are given by $\lvert a_{p}\rvert^2$ and  $\lvert a_{s}\rvert^2$, respectively. The above quantum efficiency gives the probability of a signal photon at the cavity input being converted to its sum-frequency and appear at the cavity output. It has accounted for the signal coupling loss and the SF extraction loss. It is thus a direct measure of the device figure of merit that dictates its performance in applications. 

Another measure, of less practical relevance but an intrinsic property, is the internal Rabi rotation angle $\theta=2g Q_{p,i}/\omega_p$. The system $\hat{H}_\textrm{eff}  = \hbar\mathnormal{g}(\hat{a}_p\hat{a}_s\hat{a}_{f}^\dagger+\hat{a}_p^\dagger\hat{a}_s^\dagger\hat{a}_{f})$ can be interpreted as the pump induced Rabi oscillation between signal and SF photons. $\theta$ then gives how much Rabi rotation can a pump photon induce during it is lost in the cavity. 

We implement time split-step method to numerically solve the coupled-mode equations. The parameters used in the simulation for Fig.\ref{fig3} and Fig.\ref{fig4} is listed in Table.\ref{table_s}:
\begin{table*}[ht]
\centering
\begin{tabular}{|l|l|l|l|}
\hline
Parameters & Description & Value& Unit \\
\hline
$\kappa_{p,t}$ & total dissipation rate of pump mode & $1467.9$& MHz ($\times 2\pi$)\\
\hline
$\kappa_{p,0}$ & intrinsic dissipation rate of pump mode & $270.8$& MHz ($\times 2\pi$)\\
\hline
$\kappa_{p,c}$ & coupling dissipation rate of pump mode & $1197.0$& MHz ($\times 2\pi$)\\
\hline
$\kappa_{s,t}$ & total dissipation rate of signal mode & $1510.3$& MHz ($\times 2\pi$)\\
\hline
$\kappa_{s,0}$ & intrinsic dissipation rate of signal mode & $304.4$& MHz ($\times 2\pi$)\\
\hline
$\kappa_{s,c}$ & coupling dissipation rate of signal mode & $1205.9$& MHz ($\times 2\pi$)\\
\hline
$\kappa_{f,t}$ & total dissipation rate of sum-frequency mode & $1512.2$& MHz ($\times 2\pi$)\\
\hline
$\kappa_{f,0}$ & intrinsic dissipation rate of sum-frequency mode & $318.7$& MHz ($\times 2\pi$)\\
\hline
$\kappa_{f,c}$ & coupling dissipation rate of sum-frequency mode & $1193.5$& MHz ($\times 2\pi$)\\
\hline
$d^a_{eff}$ & effective nonlinear susceptibility & 16.4& pm/V \\
\hline
$d^b_{eff}$ & effective nonlinear susceptibility & 18.2& pm/V\\
\hline
$A_{eff}$ & effective mode area & 1.0 &$\mu$m$^2$ \\
\hline
$R$ & radius of the microring & 80 &$\mu$m \\
\hline
$\xi$ & mode overlapping factor & 90\%& NA \\
\hline
\end{tabular}
\caption{Simulation parameters used in coupled-mode equations models for sum-frequency generation. $^a$: used in Fig.\ref{fig3}, $^b$: used in Fig.\ref{fig4}. The fitted $g^a$ and $g^b$ are 8.2 MHz and 9.1 MHz, respectively. }
\label{table_s}
\end{table*}
\newpage
\subsection{Experimental setup}\label{supp2}
\begin{figure}[ht]
\centering
\includegraphics[width=6 in]{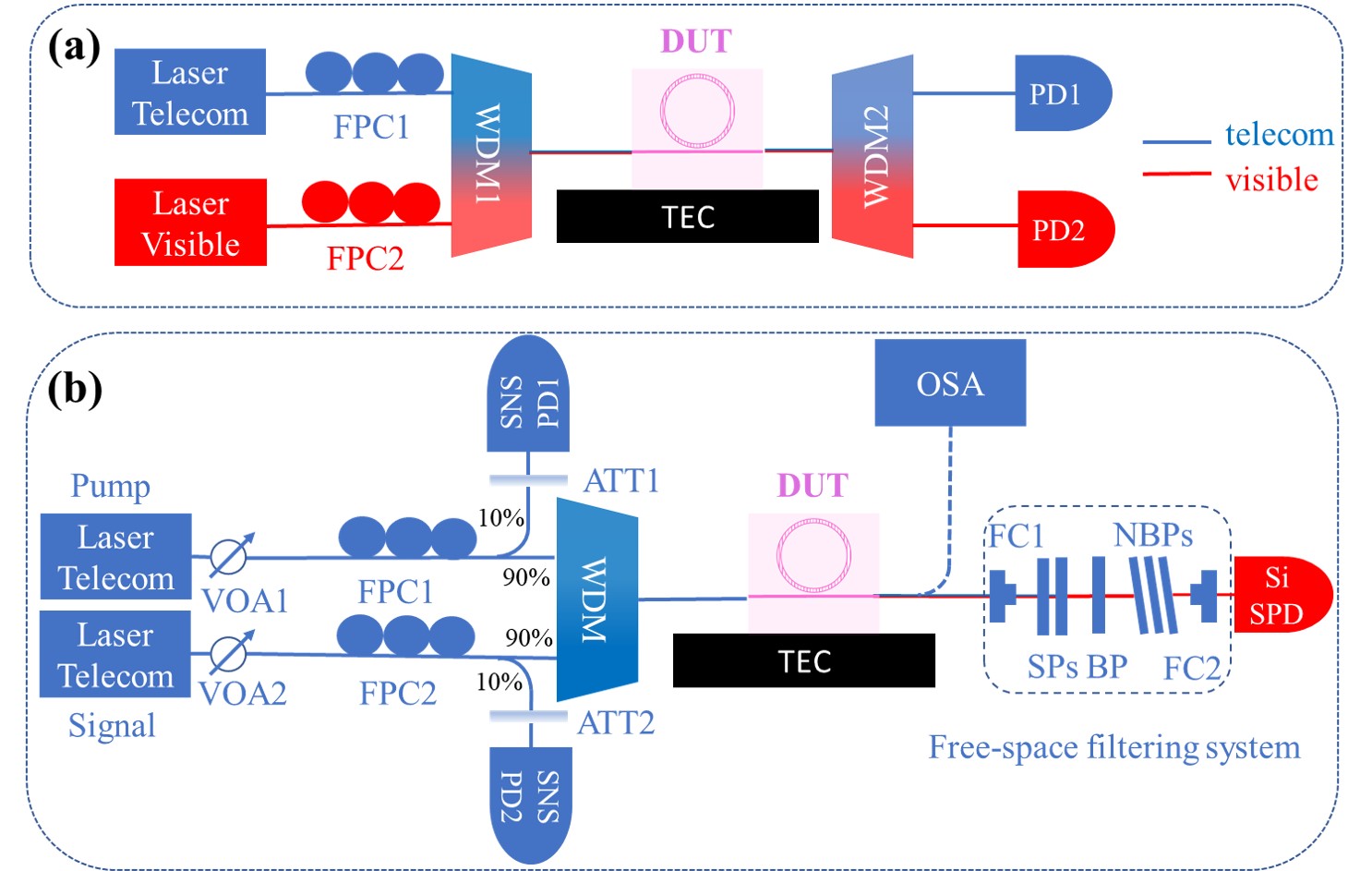}
\caption{Setups for classical (a) and quantum (b) experiments. The blue and red lines denote the telecome light path and visible path, respectively. FPC: Fiber Polarization Controller; TEC: thermoelectric cooler; WDM: wavelength-division multiplexing module; PD: photodetector; DUT: device under test; VOA:variable optical attenuator; ATT: optical attenuator; OSA: optical spectrum analyzer; FC: fiber collimator; SP: short-pass filter; BP: band-pass filter; NBP: narrow band-pass filter; SNSPD: superconducting nanowire single-photon detector; Si-SPD: silicon single-photon detector.}
\label{fig6}
\end{figure}

We use the setup shown in Fig.\ref{fig6} to characterize the device and perform the  photon conversion and interaction experiment. The whole chip is placed on a thermoelectric cooler and the temperature is set at about 45 $^\circ$C. For linear characterization, as shown in Fig.\ref{fig6}(a), we use two polarized  tunable continuous-wave (CW) lasers (Santec 550 and Newport TLB-6712) and tapered fibers (OZ OPTICS) to independently characterize the fiber-chip-fiber coupling, whose losses are measured to be (8 $\pm$ 0.15) dB around 1556 nm and (9.5 $\pm$ 0.2) dB around 778 nm, respectively. For its nonlinear optical properties, we sweep the infrared laser across the whole C-band while optimizing the chip's temperature to achieve strong second-harmonic generation (SHG). 

Once identifying the quasi-phase matching resonances, we switch to the setup in Fig.\ref{fig6}(b) for sum-frequency generation. Additional tunable CW laser (Coherent, MTP-1000) is used to serve as the signal. We will further optimize the system around its optimum SHG condition to maximize the SFG.
\begin{itemize}
\item For classical measurement, optical spectrum analyzer (OSA, Yokogawa AQ6370D) is used to collect the data. 
\item For quantum noise measurement, a silicon-based single-photon detector (Excelitas, efficiency 50$\%$, dark count 250 Hz) and a free-space filtering system are introduced. It consist of two fiber collimators (insertion loss, IL$\sim$ 2 dB), one short-pass filter ( IL$\sim$0.5 dB, extinction ratio, ER $\sim$50 dB) and one band-pass filter (10nm, 780 nm, IL$\sim$0.5 dB, ER $\sim$50 dB) rejecting pump and signal in C-band while passing through visible light, and three narrow band-pass filters (Alluxa, 3 nm, IL$\sim$1 dB, ER$>$120 dB) rejecting residue second-harmonic light of 780 nm while passing through the target light of 778 nm.
\item For interaction between single-photon coherent states, superconducting nanowire single-photon detector (ID Quantique, ID281, efficiency 85$\%$, dark count 100 Hz) are used to monitor the input photon flux. Due to limited photon-count saturation rate ($\sim$20 MHz) of SNSPDs, each monitor channel has about 50 dB attenuation. Meanwhile, the free-space filtering system will be removed to reduce the insertion loss (IL$\sim$4 dB). The SF photons will be directly measured by silicon-based single-photon detector via a lensed fiber.
\end{itemize}

\begin{thebibliography}{10}
\newcommand{\enquote}[1]{``#1''}

\bibitem{maring2018quantum}
N.~Maring, D.~Lago-Rivera, A.~Lenhard, G.~Heinze, and H.~de~Riedmatten,
  \enquote{Quantum frequency conversion of memory-compatible single photons
  from 606 nm to the telecom c-band,} {{Optica}}
  \textbf{5}, 507--513 (2018).

\bibitem{sangouard2011faithful}
N.~Sangouard, B.~Sanguinetti, N.~Curtz, N.~Gisin, R.~Thew, and H.~Zbinden,
  \enquote{Faithful entanglement swapping based on sum-frequency generation,}
  {{Physical review letters}} \textbf{106}, 120403 (2011).

\bibitem{li2019multiuser}
Y.~Li, Y.~Huang, T.~Xiang, Y.~Nie, M.~Sang, L.~Yuan, and X.~Chen,
  \enquote{Multiuser time-energy entanglement swapping based on dense
  wavelength division multiplexed and sum-frequency generation,}
  {{Physical review letters}} \textbf{123}, 250505 (2019).

\bibitem{wagenknecht2010experimental}
C.~Wagenknecht, C.-M. Li, A.~Reingruber, X.-H. Bao, A.~Goebel, Y.-A. Chen,
  Q.~Zhang, K.~Chen, and J.-W. Pan, \enquote{Experimental demonstration of a
  heralded entanglement source,} {{Nature Photonics}}
  \textbf{4}, 549--552 (2010).

\bibitem{barz2010heralded}
S.~Barz, G.~Cronenberg, A.~Zeilinger, and P.~Walther, \enquote{Heralded
  generation of entangled photon pairs,} {{Nature
  photonics}} \textbf{4}, 553--556 (2010).

\bibitem{chang2014quantum}
D.~E. Chang, V.~Vuleti{\'c}, and M.~D. Lukin, \enquote{Quantum nonlinear
  optics—photon by photon,} {{Nature Photonics}}
  \textbf{8}, 685--694 (2014).

\bibitem{li2016efficient}
Q.~Li, M.~Davan{\c{c}}o, and K.~Srinivasan, \enquote{Efficient and low-noise
  single-photon-level frequency conversion interfaces using silicon
  nanophotonics,} {{Nature Photonics}} \textbf{10},
  406--414 (2016).

\bibitem{lu2019efficient}
X.~Lu, G.~Moille, Q.~Li, D.~A. Westly, A.~Singh, A.~Rao, S.-P. Yu, T.~C.
  Briles, S.~B. Papp, and K.~Srinivasan, \enquote{Efficient telecom-to-visible
  spectral translation through ultralow power nonlinear nanophotonics,}
  {{Nature Photonics}} \textbf{13}, 593--601 (2019).

\bibitem{wang2021efficient}
J.-Q. Wang, Y.-H. Yang, M.~Li, X.-X. Hu, J.~B. Surya, X.-B. Xu, C.-H. Dong,
  G.-C. Guo, H.~X. Tang, and C.-L. Zou, \enquote{Efficient frequency conversion
  in a degenerate $\chi$ (2) microresonator,} {{Physical
  Review Letters}} \textbf{126}, 133601 (2021).

\bibitem{chang2019strong}
L.~Chang, A.~Boes, P.~Pintus, J.~D. Peters, M.~Kennedy, X.-W. Guo, N.~Volet,
  S.-P. Yu, S.~B. Papp, and J.~E. Bowers, \enquote{Strong frequency conversion
  in heterogeneously integrated gaas resonators,} {{APL
  Photonics}} \textbf{4}, 036103 (2019).

\bibitem{wilson2020integrated}
D.~J. Wilson, K.~Schneider, S.~H{\"o}nl, M.~Anderson, Y.~Baumgartner,
  L.~Czornomaz, T.~J. Kippenberg, and P.~Seidler, \enquote{Integrated gallium
  phosphide nonlinear photonics,} {{Nature Photonics}}
  \textbf{14}, 57--62 (2020).

\bibitem{chang2020ultra}
L.~Chang, W.~Xie, H.~Shu, Q.-F. Yang, B.~Shen, A.~Boes, J.~D. Peters, W.~Jin,
  C.~Xiang, S.~Liu \emph{et~al.}, \enquote{Ultra-efficient frequency comb
  generation in algaas-on-insulator microresonators,}
  {{Nature communications}} \textbf{11}, 1--8 (2020).

\bibitem{ma2020ultrabright}
Z.~Ma, J.-Y. Chen, Z.~Li, C.~Tang, Y.~M. Sua, H.~Fan, and Y.-P. Huang,
  \enquote{Ultrabright quantum photon sources on chip,}
  {{Physical Review Letters}} \textbf{125}, 263602 (2020).

\bibitem{chen2021efficient}
J.-Y. Chen, C.~Tang, M.~Jin, Z.~Li, Z.~Ma, H.~Fan, S.~Kumar, Y.~M. Sua, and
  Y.-P. Huang, \enquote{Efficient frequency doubling with active stabilization
  on chip,}  (2021).

\bibitem{lu2019periodically}
J.~Lu, J.~B. Surya, X.~Liu, A.~W. Bruch, Z.~Gong, Y.~Xu, and H.~X. Tang,
  \enquote{Periodically poled thin-film lithium niobate microring resonators
  with a second-harmonic generation efficiency of 250,000\%/w,}
  {{Optica}} \textbf{6}, 1455--1460 (2019).

\bibitem{chen2019ultra}
J.-Y. Chen, Z.-H. Ma, Y.~M. Sua, Z.~Li, C.~Tang, and Y.-P. Huang,
  \enquote{Ultra-efficient frequency conversion in quasi-phase-matched lithium
  niobate microrings,} {{Optica}} \textbf{6}, 1244--1245
  (2019).

\bibitem{Lu:20}
J.~Lu, M.~Li, C.-L. Zou, A.~A. Sayem, and H.~X. Tang, \enquote{Toward 1\%
  single-photon anharmonicity with periodically poled lithium niobate microring
  resonators,} {{Optica}} \textbf{7}, 1654--1659 (2020).

\bibitem{gao2021broadband}
R.~Gao, H.~Zhang, F.~Bo, W.~Fang, Z.~Hao, N.~Yao, J.~Lin, J.~Guan, L.~Deng,
  M.~Wang, L.~Qiao, and Y.~Cheng, \enquote{Broadband highly efficient nonlinear
  optical processes in on-chip integrated lithium niobate microdisk resonators
  of q-factor above 10$^8$,}  (2021).

\bibitem{ilchenko2004nonlinear}
V.~S. Ilchenko, A.~A. Savchenkov, A.~B. Matsko, and L.~Maleki,
  \enquote{Nonlinear optics and crystalline whispering gallery mode cavities,}
  {{Physical review letters}} \textbf{92}, 043903 (2004).

\bibitem{li2017nonlinear}
Y.~Li, T.~Xiang, Y.~Nie, M.~Sang, and X.~Chen, \enquote{Nonlinear interaction
  between broadband single-photon-level coherent states,}
  {{Photonics Research}} \textbf{5}, 324--328 (2017).

\bibitem{breunig2016three}
I.~Breunig, \enquote{Three-wave mixing in whispering gallery resonators,}
  {{Laser \& Photonics Reviews}} \textbf{10}, 569--587
  (2016).

\bibitem{hosseini2010systematic}
E.~S. Hosseini, S.~Yegnanarayanan, A.~H. Atabaki, M.~Soltani, and A.~Adibi,
  \enquote{Systematic design and fabrication of high-q single-mode
  pulley-coupled planar silicon nitride microdisk resonators at visible
  wavelengths,} {{Optics express}} \textbf{18}, 2127--2136
  (2010).

\bibitem{fan2021photon}
H.~Fan, Z.~Ma, J.~Chen, Z.~Li, C.~Tang, Y.~Sua, and Y.~Huang, \enquote{Photon
  conversion in thin-film lithium niobate nanowaveguides: A noise analysis,}
  {{arXiv preprint arXiv:2102.07044}}  (2021).

\bibitem{singh2019quantum}
A.~Singh, Q.~Li, S.~Liu, Y.~Yu, X.~Lu, C.~Schneider, S.~H{\"o}fling, J.~Lawall,
  V.~Verma, R.~Mirin \emph{et~al.}, \enquote{Quantum frequency conversion of a
  quantum dot single-photon source on a nanophotonic chip,}
  {{Optica}} \textbf{6}, 563--569 (2019).

\bibitem{vazirani2019fully}
U.~Vazirani and T.~Vidick, \enquote{Fully device independent quantum key
  distribution,} {{Communications of the ACM}}
  \textbf{62}, 133--133 (2019).

\bibitem{Chen:19}
J.~yang Chen, Y.~M. Sua, Z.~hui Ma, C.~Tang, Z.~Li, and Y.~ping Huang,
  \enquote{Efficient parametric frequency conversion in lithium niobate
  nanophotonic chips,} {{OSA Continuum}} \textbf{2},
  2914--2924 (2019).

\bibitem{chen2020efficient}
J.-Y. Chen, C.~Tang, Z.-H. Ma, Z.~Li, Y.~M. Sua, and Y.-P. Huang,
  \enquote{Efficient and highly tunable second-harmonic generation in z-cut
  periodically poled lithium niobate nanowaveguides,}
  {{Optics Letters}} \textbf{45}, 3789--3792 (2020).

\bibitem{jin2019high}
M.~Jin, J.-Y. Chen, Y.~M. Sua, and Y.-P. Huang, \enquote{High-extinction
  electro-optic modulation on lithium niobate thin film,}
  {{Optics letters}} \textbf{44}, 1265--1268 (2019).

\bibitem{he2019self}
Y.~He, Q.-F. Yang, J.~Ling, R.~Luo, H.~Liang, M.~Li, B.~Shen, H.~Wang,
  K.~Vahala, and Q.~Lin, \enquote{Self-starting bi-chromatic linbo 3 soliton
  microcomb,} {{Optica}} \textbf{6}, 1138--1144 (2019).

\bibitem{wang2019monolithic}
C.~Wang, M.~Zhang, M.~Yu, R.~Zhu, H.~Hu, and M.~Loncar, \enquote{Monolithic
  lithium niobate photonic circuits for kerr frequency comb generation and
  modulation,} {{Nature communications}} \textbf{10}, 1--6
  (2019).

\bibitem{zhang2019broadband}
M.~Zhang, B.~Buscaino, C.~Wang, A.~Shams-Ansari, C.~Reimer, R.~Zhu, J.~M. Kahn,
  and M.~Lon{\v{c}}ar, \enquote{Broadband electro-optic frequency comb
  generation in a lithium niobate microring resonator,}
  {{Nature}} \textbf{568}, 373--377 (2019).

\bibitem{gong2020near}
Z.~Gong, X.~Liu, Y.~Xu, and H.~X. Tang, \enquote{Near-octave lithium niobate
  soliton microcomb,} {{Optica}} \textbf{7}, 1275--1278
  (2020).

\bibitem{jiuyi}
J.~Zhang, Y.~M. Sua, J.-Y. Chen, J.~Ramanathan, C.~Tang, Z.~Li, Y.~Hu, and
  Y.-P. Huang, \enquote{Carbon-dioxide absorption spectroscopy with solar
  photon counting and integrated lithium niobate micro-ring resonator,}
  {{Applied Physics Letters}} \textbf{118}, 171103 (2021).

\bibitem{boes2018status}
A.~Boes, B.~Corcoran, L.~Chang, J.~Bowers, and A.~Mitchell, \enquote{Status and
  potential of lithium niobate on insulator (lnoi) for photonic integrated
  circuits,} {{Laser \& Photonics Reviews}} \textbf{12},
  1700256 (2018).

\bibitem{zhang2008demonstration}
Q.~Zhang, X.-H. Bao, C.-Y. Lu, X.-Q. Zhou, T.~Yang, T.~Rudolph, and J.-W. Pan,
  \enquote{Demonstration of a scheme for the generation of “event-ready”
  entangled photon pairs from a single-photon source,}
  {{Physical Review A}} \textbf{77}, 062316 (2008).


\end{thebibliography}
\end{document}